\documentclass[a4paper,12pt]{article}
\pdfoutput=1
\usepackage[margin=1in]{geometry}
\usepackage{graphicx,xcolor}
\usepackage{mathtools,braket,blkarray}
\allowdisplaybreaks
\usepackage{amssymb,amsfonts,bbm,relsize}
\usepackage[width=0.9\textwidth,footnotesize]{caption}
\usepackage[font=scriptsize]{subcaption}
\usepackage{cite,hyperref,url}
\hypersetup{pageanchor=false}
\usepackage{booktabs,multirow,makecell}
\usepackage[utf8]{inputenc}
\usepackage{cleveref}
\usepackage{amsmath}

\begin{document}

\begin{center}
{\bf\Large Emergent gravity from off-shell energy fixing
  } \\[12mm]
Francisco~J.~de~Anda$^{\star}$%
\footnote{E-mail: \texttt{fran@tepaits.mx}},
\\[-2mm]

\end{center}
\vspace*{0.50cm}
\centerline{$^{\star}$ \it
Tepatitl{\'a}n's Institute for Theoretical Studies, C.P. 47600, Jalisco, M{\'e}xico}
\vspace*{1.20cm}

\begin{abstract}
{\noindent
Off-shell processes do not preserve the Energy Momentum Tensor (EMT) in QFT. Fixing the EMT throughout off-shell processes, implies a graviton-like quantum field to emerge without dynamics. Its dynamics are generated through quantum corrections. This Fixed Off-Shell Energy Condition (FOSEC)  implies the existence of a linear gravity-like theory, and in special cases the full Poincar\`e gauge theory. In this work it is shown that imposing the FOSEC in QFT implies the emergence of a viable quantum theory of gravity.
}
\end{abstract}

\section*{Introduction}

The quantization of gravity remains as one of the most important problems to be solved in theoretical physics. 
The fact that the standard quantization approach does not work for gravity, suggests that its nature is different from other forces. One approach is to treat gravity as an emergent phenomenon instead of a fundamental one \cite{Verlinde:2010hp,Sindoni:2011ej,Linnemann:2017hdo,Barcelo:2001ah,Padmanabhan:2016bha}. In this approach, the gravitational degrees of freedom emerge from the dynamics of fundamental ones.

The emergence of gravity can happen at different levels \cite{Oriti:2018dsg}.
Since the standard theory for gravity is General Relativity (GR), one can search for the emergence of the spacetime curvature from flat spacetime processes. Once the curvature appears, its dynamics can be induced through quantum corrections \`a la Sakharov \cite{Sakharov:1967pk,Barcelo:2001tb}. Any acceptable theory should resemble GR in some limit \cite{Capozziello:2011et,Nojiri:2010wj}.

In GR, the source of spacetime curvature is the Energy Momentum Tensor (EMT), which exists regardless of the existence of gravity in the theory. This makes the EMT the natural source for the emergence of gravity \cite{Carone:2016tup}.

A system that has translation invariance, has a conserved EMT. However off-shell field configurations do not conserve it, even with translation invariance. This is usually disregarded since off-shell processes are, by definition, non observable.

The aim of the paper is to explore the consequences of fixing the EMT through off-shell processes. 
It will be shown that imposing the Fixed Off-Shell Energy Condition (FOSEC) is equivalent to adding a graviton field without dynamics from which a full theory of gravity can be induced. 

The outline of the paper is as follows: in section \ref{sec:fosec}, the implications of the FOSEC are studied. It is shown that the FOSEC is equivalent to adding a graviton without dynamics. In section \ref{sec:id} the graviton dynamics are induced \`a la Sakharov. In section \ref{sec:poincare}, it is shown that in special cases, the FOSEC can generate the full Poincar\`e gauge theory, which contains GR as a subset. In section \ref{sec:eg} it is discussed the equivalence of the FOSEC complying theory to an emergent gravitational theory and how it circumvents some of the usual problems of quantum gravity.

\section{Fixed off-shell EMT}
\label{sec:fosec}

The EMT is conserved due to the translation invariance of a lagrangian 
\begin{equation}
T_{\mu\nu}(\Psi)=\frac{\delta\mathcal{L}}{\delta\partial^\mu \Psi}\partial_\nu\Psi-\eta_{\mu\nu}\mathcal{L},
\end{equation}
where $\Psi$ represents an arbitrary field, and the lagrangian $\mathcal{L}$ is a functional of $\Psi$. This is the canonical definition of the EMT, which is not necessarily symmetric nor gauge invariant. It can be made so by adding an identically conserved superpotential term\cite{Blaschke:2016ohs}
\begin{equation}
T_{\mu\nu}\to T_{\mu\nu}+\partial^\rho \chi_{\mu\rho\nu},\ \ \ {\rm with}\ \ \ \chi_{\mu\rho\nu}=-\chi_{\rho\mu\nu}.
\end{equation}

The observable field configurations, called on-shell, comply with their corresponding equations of motion. The field configurations that do not comply with them are the non observable virtual particles called off-shell field configurations. The EMT is conserved
\begin{equation}
\partial^\mu T_{\mu\nu}(\Psi)=\partial_\nu \Psi\left(\partial_\mu\frac{\delta\mathcal{L}}{\delta\partial_\mu \Psi} -\frac{\delta\mathcal{L}}{\delta\Psi} \right),
\end{equation}
when the equations of motion are fulfilled i.e. when the term inside ellipsis vanish. Therefore off-shell field configurations do not conserve the EMT. This is not a problem since they are not observable. It would be interesting to build a theory that conserves the EMT even through off-shell processes.

QFT can be treated through the path integral formalism. It based on the lagrangian $\mathcal{L}(\Psi)$ of a set of classical fields $\Psi$. Instead of obtaining the physical observables from the lagrangian directly, they are obtained from the partition function
\begin{equation}
Z=\int \mathcal{D}\Psi e^{i S(\Psi)}=\int \mathcal{D}\Psi e^{i\int d^4x\ \mathcal{L}(\Psi)},
\end{equation}
which is a functional integral over all possible field (on and off-shell) configurations. This partition function $Z$ is infinite and should be normalized. All physical observables $\mathcal{O}$ can be obtained from this partition function as \cite{Weinberg:1995mt, Weinberg:1996kr} 
\begin{equation}
\braket{\mathcal{O}(\Psi)}=\frac{\int \mathcal{D}\Psi\ \mathcal{O}(\Psi)\ e^{i\int d^4x\ \mathcal{L}(\Psi)}}{\int \mathcal{D}\Psi e^{i\int d^4x\ \mathcal{L}(\Psi)}}.
\end{equation}
The path integral method is the most symmetrical way to treat QFT \cite{Cartier:2010zqa,Glimm:1987ng,rivers}.
In this formalism one can impose the FOSEC by limiting the integration space to only the to only the field configurations that fulfill the FOSEC. This is defined as
\begin{equation}
Z=\int_{T_{\mu\nu}(\Psi)-\mathcal{T}_{\mu\nu}^0} \mathcal{D}\Psi e^{i\int d^4x\ \mathcal{L}(\Psi)},
\end{equation}
where $T_{\mu\nu}(\Psi)$ is the functional EMT of the fields $\Psi$ while $\mathcal{T}_{\mu\nu}^0$ is a function defining the total energy-momentum of the specific considered system.
One can limit the range of the integration by allowing the integration over all field configurations but introducing a $\delta$ Dirac measure
\begin{equation}
Z'=\int \mathcal{D}\Psi e^{i\int d^4x\ \mathcal{L}(\Psi)}\ \delta(T_{\mu\nu}(\Psi)-\mathcal{T}_{\mu\nu}^0).
\label{eq:zprime}
\end{equation}
The addition of this Dirac measure to the functional integral defines the FOSEC. This way the functional integral is evaluated only through field configurations that have their EMT fixed to be $\mathcal{T}^{0}_{\mu\nu}$. 

The introduction of the $\delta$ Dirac measure in the functional integral reduces the degrees of freedom of the fields.
It is clear that the FOSEC can only be applied on systems with multiple fields so that the functional integral is evaluated over more than 6 degrees of freedom (since the EMT has 6 independent components). In theories with less than that, the functional integral either could be evaluated exactly, therefore suppressing quantum effects, or it couldn't be evaluated at all. 
In realistic models, such as the Standard Model, the number of fields is much larger, so the FOSEC can be applied consistently. The FOSEC fixes the total EMT but the energy or momentum of a specific field is never known exactly, complying the uncertainty principle. 

The effects of adding a total derivative to the lagrangian are usually ignored as they don't affect the equations of motion. However they affect the EMT since it has a term proportional to the lagrangian.  A total derivative in the lagrangian has physical effects when gravity is present. The $\mathcal{T}^0_{\mu\nu}$ must the total energy of the system, including the energy of the vacuum (which is a total derivative) to impose the FOSEC consistently. 

Quantum fluctuations define the vacuum energy. Specifically the vacuum bubble diagrams define a constant energy everywhere, which is infinite. Since it is only a constant, it won't have any physical impact in the theory and can be renormalized to zero.
One could assume a non zero value for the vacuum energy and interpret  that quantum processes draw energy from there. 
The physical motivation of the FOSEC is to limit the quantum fluctuations of the fields so that they don't draw more than the available energy from the vacuum. The energy that the quantum processes need, must come from the vacuum energy, which is assumed finite in this work. 

The definition of the FOSEC in eq. \ref{eq:zprime}, needs a $\delta$ Dirac measure. 
 Using the functional integral (up to irrelevant constants \cite{horma})
\begin{equation}
\delta(\psi-\phi)=\int\mathcal{D}\xi\ e^{-i\int d^4 x\ \xi(\psi-\phi)},
\end{equation}
 the primed partition function can be rewritten as
\begin{equation} \begin{split}
Z'=\int \mathcal{D}\Psi\  e^{i\int d^4x\ \mathcal{L}(\Psi)}\ \delta(T_{\mu\nu}(\Psi)-\mathcal{T}_{\mu\nu}^0)=\int \mathcal{D}\Psi\mathcal{D}h\  e^{i\int d^4x\ \mathcal{L}'(\Psi,h)},\\
{\rm with}\ \ \ \mathcal{L}'(\Psi,h)=\mathcal{L}(\Psi)-\frac{1}{2}h^{\mu\nu}\left(T_{\mu\nu}(\Psi)-\mathcal{T}^0_{\mu\nu}\right),
\label{eq:genlag}
\end{split}\end{equation}
where there is an arbitrary coupling constant which can be reabsorbed into the $h^{\mu\nu}$.
The $Z'$ can be rewritten in the standard form
\begin{equation}
Z'=\int \mathcal{D}\tilde{\Psi}\  e^{i\int d^4x\ \mathcal{L}'(\tilde{\Psi})},
\end{equation}
where $\tilde{\Psi}=(\Psi,h)$, fulfilling all the necessary condition for a consistent probability measure \cite{Glimm:1987ng}. One could arrive at the same result by having a standard partition function and adding a field $h^{\mu\nu}$ to the lagrangian as a Lagrange multiplier to fulfill the FOSEC. This is the standard path integral formalism and all the changes amount to the new interaction to the added field $h^{\mu\nu}$.
This field and the $\mathcal{T}^0_{\mu\nu}$ must be invariant under all internal symmetries, as the EMT is. They must transform as a rank 2 tensor under Lorentz symmetry, so that all the original symmetries are preserved. If $\mathcal{T}^0_{\mu\nu}$ is required to be Lorentz invariant, it must have the form $\mathcal{T}^0_{\mu\nu}=\Lambda_0 \eta_{\mu\nu}$. If $\mathcal{T}^0_{\mu\nu}$ is required to be scale invariant, it must vanish.

The FOSEC is equivalent to adding a quantum field $h^{\mu\nu}$ coupling to the original fields through their EMT, just as a graviton would. There is a functional integration over $h^{\mu\nu}$, as there is for any quantized field. There are no dynamical terms for this graviton-like field.

The coupling of the emergent graviton to matter is not the standard one, since it couples to the difference $T_{\mu\nu}-\mathcal{T}^0_{\mu\nu}$. This fact  important cosmological implications. One can impose the FOSEC by requiring the total EMT with vacuum contributions is always a constant, i.e. that there is a fixed positive amount of energy everywhere. Therefore when a particle is created, the energy of the vacuum is lowered so that the sum of the energy of the particle and the vacuum is always constant. This is done by fixing
 \begin{equation}
\mathcal{T}^0_{\mu\nu}=\Lambda_0\eta_{\mu\nu}.
\end{equation}
In this case the coupling of the graviton is the standard one, to the EMT of the fields, with $\Lambda_0$ a negative cosmological constant.
 
\subsection{EMT fixing in QFT: Scalar fields}

The lagrangian of a complex scalar field $\phi$ in flat spacetime can be written as
\begin{equation}
\mathcal{L}(\phi)=\eta^{\mu\nu}\left(\partial_\mu \phi\right)^\dagger \partial_\nu \phi-m^2\phi\phi^\dagger,
\end{equation}
where $\eta^{\mu\nu}={\rm diag}(1,-1,-1,-1)$ is the Minkowski metric and $m$ is its mass.
The symmetric EMT of the complex scalar field is 
\begin{equation}
T_{\mu\nu}(\phi,\phi^\dagger)=2\left(\partial_\mu\phi\right)^\dagger \partial_\nu\phi-\eta_{\mu\nu}\eta^{\alpha\beta}\left(\partial_\alpha\phi\right)^\dagger \partial_\beta\phi+\eta_{\mu\nu} m^2\phi\phi^\dagger.
\end{equation}
The FOSEC effective lagrangian is
\begin{equation}\begin{split}
\mathcal{L}'(\phi,h)&=\left(\eta^{\mu\nu}\left(1+\frac{1}{2}h\right)-h^{\mu\nu}\right)\left(\partial_\mu\phi\right)^\dagger \partial_\nu\phi-\left(1+\frac{1}{2}h\right)m^2\phi\phi^\dagger+\frac{1}{2}h^{\mu\nu}\mathcal{T}^0_{\mu\nu},
\label{eq:fosecsc}
\end{split}\end{equation}
where $h=h^{\mu\nu}\eta_{\mu\nu}$.
By defining the metric
\begin{equation}
g_{\mu\nu}=\eta_{\mu\nu}+h_{\mu\nu},
\label{eq:gmn}
\end{equation}
and keeping only the linear terms, one obtains
\begin{equation}
g^{\mu\nu}=\eta^{\mu\nu}-h^{\mu\nu}+\mathcal{O}(h^2),\ \ \ \sqrt{-g}=1+\frac{1}{2}h+\mathcal{O}(h^2),
\label{eq:gh}
\end{equation}
with $g=\det(g^{\mu\nu})$.
The effective lagrangian can be written as
\begin{equation}
\mathcal{L}''(\phi,g)=\sqrt{-g}\left(g^{\mu\nu}\left(\partial_\mu\phi\right)^\dagger \partial_\nu\phi-m^2\phi\phi^\dagger-\frac{1}{2}g^{\mu\nu}\mathcal{T}_{\mu\nu}^0+\frac{1}{2}\mathcal{T}^0\right)=\mathcal{L}'(\phi,h)+\mathcal{O}(h^2).
\end{equation}
One concludes that the effective lagrangian of a scalar field complying with the FOSEC, is equivalent to the scalar field lagrangian in a curved spacetime (with a source term $\mathcal{T}_{\mu\nu}^0$) up to linear order.

\subsection{EMT fixing in QFT: Fermionic fields}
\label{sec:fermions}

The lagrangian of a spin $1/2$ fermion $\psi$ is
\begin{equation}
\mathcal{L}(\psi)=\frac{i}{2}\bar{\psi}\gamma^\mu\partial_\mu\psi-\frac{i}{2}\partial_\mu\bar{\psi}\gamma^\mu\psi-m\bar\psi\psi,
\end{equation}
with a canonical EMT \cite{Gu:2006eu}
\begin{equation}
T_{\mu\nu}(\psi)=\frac{i}{2}\left(\bar{\psi}\gamma_{(\mu}\partial_{\nu)}\psi-\partial_{(\mu}\bar{\psi}\gamma_{\nu)}\psi\right)-\frac{i}{2}\eta_{\mu\nu}\left(\bar{\psi}\gamma^\mu\partial_\mu\psi-\partial_\mu\bar{\psi}\gamma^\mu\psi\right)+m\eta_{\mu\nu}\bar\psi\psi.
\end{equation}
The FOSEC effective lagrangian is
\begin{equation}
\mathcal{L}'(\psi,h)=\frac{i}{2}\left(\eta^{\mu\nu}\left(1+h\right)-2h^{\mu\nu}\right)\left(\bar{\psi}\gamma_\mu\partial_\nu\psi-\partial_\nu\bar\psi \gamma_\mu \psi\right)-m(1+h)\bar\psi \psi+\mathcal{T}_{\mu\nu}^0h^{\mu\nu}.
\label{eq:fosecdir}
\end{equation}
As done before, this new lagrangian coincides with the one of linearized gravity with no dynamic term for the graviton.
The usual curved lagrangian for a fermion uses a first order formalism, using the vierbein instead of the metric 
\begin{equation}e^a_\mu e^b_\nu\eta_{ab}=g_{\mu\nu},\end{equation} and it requiress the spin connection $\Omega_\mu$.
The curved spacetime lagrangian for a fermion (with a source $\mathcal{T}^0_{\mu\nu}$ terms) is
\begin{equation}
\mathcal{L}''(\psi,e)=\frac{i}{2}|e|\left(\bar{\psi}e^\mu_a\gamma^a\left(\partial_\mu-\Omega_\mu\right)\psi\right)-m|e|\bar\psi\psi-\frac{|e|}{2} e_a^\mu e_b^\nu\eta^{ab}\mathcal{T}_{\mu\nu}^0+\frac{|e|}{2}\mathcal{T}^0+h.c.
\label{eq:curvfer}
\end{equation}
Linearizing the vierbein as
\begin{equation}
g_{\mu\nu}=\eta_{\mu\nu}+h_{\mu\nu},\ \ \ e^a_\mu=\delta^a_\mu+\frac{1}{2}h^a_\mu+\mathcal{O}(h^2),\ \ \ e_a^\mu=\delta_a^\mu-\frac{1}{2}h_a^\mu+\mathcal{O}(h^2),
\end{equation} 
the expansion of the connection is $\Omega=\mathcal{O}(h^2)$, so that it doesn't contribute at linear order.  Therefore the equivalence  holds \cite{Aldrovandi:1992di}
\begin{equation}
\mathcal{L}''(\psi,\delta+h)=\mathcal{L}'(\psi,h)+\mathcal{O}(h^2),
\end{equation}
just as the in the previous case.

\subsection{EMT fixing in QFT: Generalization}
\label{sec:gener}

It has been shown that imposing the FOSEC is equivalent to introducing an auxiliary field $h^{\mu\nu}$ reminiscent of the graviton without dynamical terms. In general, the linearization of a lagrangian of an arbitrary field (or collection of fields) $\Psi$ in curved space is
\begin{equation}
\mathcal{L}''(\Psi,g)=\mathcal{L}(\Psi)-h^{\mu\nu}\left.\frac{\delta\mathcal{L}''(\Psi,g)}{\delta g^{\mu\nu}}\right|_{g^{\mu\nu}\to h^{\mu\nu}}+\mathcal{O}(h^2),
\label{eq:ghmunu}
\end{equation}
where the functional derivative with respect to $g^{\mu\nu}$ is the Einstein-Hilbert EMT. It is equivalent to the symmetrized canonical EMT \cite{Blaschke:2016ohs}.

The conservation of the EMT is a consequence of translation invariance of the lagrangian
\begin{equation}
x^\mu\to x^\mu+\xi^\mu,
\end{equation}
with $\xi^\mu$ an arbitrary constant vector. By adding the $h^{\mu\nu}$ field, this global symmetry becomes a local one, with the gauge transformation
\begin{equation}\begin{split}
x^\mu&\to x^\mu+\xi^\mu(x),\\ 
h^{\mu\nu}(x)&\to h^{\mu\nu}(x)+\partial^\mu\xi^\nu-\partial^\nu\xi^\mu.
\label{eq:difft}
\end{split}\end{equation}
These are known as the diffeomorphism transformations. All lagrangians obtained through the FOSEC, as in eqs. \ref{eq:genlag} and \ref{eq:ghmunu}, are invariant under this local symmetry \cite{Alvarez:2007cp,Scharf:2001bk}. 

\subsubsection{Global to local symmetry}
\label{sec:globloc}

Following Noether's theorem, a lagrangian invariant under a continuous transformation, 
\begin{equation}
x^\mu\to x^\mu+\delta x^\mu,\ \ \ \Psi\to\Psi+\delta\Psi,
\end{equation}
has the conserved current
\begin{equation}
j_\mu=\frac{\delta\mathcal{L}}{\delta \partial^\mu \Psi}\delta \Psi-\left(\frac{\delta\mathcal{L}}{\delta\partial^\mu \Psi}\partial_\nu\Psi-\eta_{\mu\nu}\mathcal{L}\right)\delta x^\nu.
\end{equation}
The second term is just the canonical EMT. Previous sections have studied the effects of fixing the EMT through off-shell processes. One can also study the effects of fixing off-shell the conserved current due to an internal symmetry.

Let the lagrangian of the fields $\Psi$,  have a global internal symmetry corresponding to a Lie group $G$, with generators $T^a$ (where $a$ runs through the algebra indices). A field in the fundamental representation transforms as
\begin{equation}
\Psi\to e^{iT^a\alpha_a}\Psi,
\end{equation}
where the $\alpha_a$ are arbitrary constants. The symmetry implies the conservation of the vector current
\begin{equation}
j_\mu^a =iT^a \Psi\frac{\delta\mathcal{L}}{\delta \partial^\mu \Psi}.
\end{equation}
If this current is required to be fixed through off-shell processes to be $\mathcal{J}_\mu^{a0}$, then
\begin{equation}\begin{split}
Z'=\int \mathcal{D}\Psi\  e^{i\int d^4x\ \mathcal{L}(\Psi)}\ \delta(j_\mu^a(\Psi)-\mathcal{J}_\mu^{a0})=\int \mathcal{D}\Psi\mathcal{D}A\  e^{i\int d^4x\ \mathcal{L}'(\Psi,A)},\\ 
{\rm with}\ \ \ \mathcal{L}'(\Psi,A)=\mathcal{L}(\Psi)-i\tilde{g} A^\mu_a \left(T^a\Psi \frac{\delta\mathcal{L}}{\delta \partial^\mu \Psi}-\mathcal{J}^{a0}_{\mu}\right),
\end{split}\end{equation}
where $\tilde{g}$ is an arbitrary constant that becomes the gauge coupling. This is equivalent to changing derivatives to covariant gauge derivatives through the minimal coupling, up to linear order. 

The $\mathcal{J}^{a0}_{\mu}$ would define the on-shell value of the current of a specific system. Note that in the special case that it vanishes, the effective lagrangian adds the minimal coupling to a gauge field up to linear order.
Therefore, starting from a lagrangian with a global symmetry and imposing the vanishing ($\mathcal{J}_\mu^{a0}=0$) of its conserved current off-shell, generates a gauge theory up to linear order in its coupling $\tilde{g}$.

One concludes that in a system with a global symmetry, the imposition of the conserved current to be fixed through off-shell processes, changes the global symmetry to a local symmetry (up to a lineal order).

\section{Induced dynamics}
\label{sec:id}

Imposing the FOSEC, effectively  adds a quantum field $h^{\mu\nu}$ to the lagrangian, with no dynamical terms. 
These arise from quantum corrections. 

\subsection{Scalar field}
In the scalar case, one can rewrite the FOSEC fulfilling action from eq. \ref{eq:fosecsc} as
\begin{equation}
S'(\phi,h)=\int d^4 x\mathcal{L}'(\phi,h)=\int d^4 x\left( \phi^\dagger D_s \phi+\frac{1}{2}h^{\mu\nu}\mathcal{T}^0_{\mu\nu}\right),
\end{equation}
with 
\begin{equation}
\begin{split}
D_s&=-\left[\eta^{\mu\nu}\left(1+\frac{1}{2}h\right)-h^{\mu\nu}\right]\partial_\mu\partial_\nu-\left[\frac{1}{2}\eta^{\mu\nu}\partial_\mu h-\partial_\mu h^{\mu\nu}\right]\partial_\nu-\left(1+\frac{1}{2}h\right)m^2.
\label{eq:ds}
\end{split}
\end{equation}
Using the standard background field method for the scalar field \cite{Adler:1982ri,Ichinose:1981uw}, the first loop contribution is
\cite{Obukhov:1983mm,Avramidi:2001ns,Vassilevich:2003xt,Avramidi:2009tn,Chaichian:2018xdo}
\begin{equation}
S'^{(1)}(h)=i\ln \det (  D_s/D^0_s)=i{\rm Tr}\ln( D_s/D^0_s),
\end{equation}
where $D^0_s$ is the operator for a suitable background reference (for example $D_s^0=D_s|_{h=0}$) and it is necessary to use the identity
\begin{equation}
S'^{(1)}(h)=i{\rm Tr}\ln( D_s/D^0_s)=-i{\rm Tr}\int_0^\infty \frac{d\tau}{\tau}\left(e^{-i\tau D_s}-e^{-i\tau D_s^0}\right).
\end{equation}
The $D_s^0$ cancels infinities, but will contribute to the final lagrangian as a constant that can be ignored. One can write the exponential operator in terms of the kernel function
\begin{equation}
e^{-i\tau D_s}\phi(x)=\int d^4 y K(\tau;x,y)\phi(y),
\end{equation}
which satisfies the Schrödinger-like equation
\begin{equation}
i\frac{d}{d\tau}K(\tau;x,y)=D_s K(\tau; x,y),\ \ \ \ K(0;x,y)=\delta(x-y).
\end{equation}
The first loop contribution can then be rewritten as (ignoring the constants coming from $D_s^0$)
\begin{equation}
S'^{(1)}(h)=-i{\rm Tr}\int_0^\infty \frac{d\tau}{\tau}e^{-i\tau D_s}=-i\int_0^\infty \frac{d\tau}{\tau}\int d^4 x\ {\rm tr}\ K(\tau; x,x).
\label{eq:tau}
\end{equation}
One can use the heat kernel expansion \cite{Avramidi:2000bm}
\begin{equation}
K(\tau,x,x)=\frac{i}{(4\pi i \tau)^2}\sum_{n=0}^\infty
(i\tau)^{n} A_{n}(x),
\end{equation}
where $A_n$ are invariants built from $D_s$.
From eq. \ref{eq:ds}, one can define the differential operator (acting on a vector $v_\nu$) as
\begin{equation}
\nabla^s_\mu v_\nu=\left(\left[\delta^\alpha_\nu\left(1+\frac{h}{2}\right)-\eta_{\beta\nu}h^{\beta\alpha}\right]\partial_\mu-\delta^\alpha_\nu\partial_\mu\frac{h}{2}+\eta_{\beta\nu}\partial_\mu h^{\beta\alpha} \right)v_\alpha,
\label{eq:nablas}
\end{equation}
and the function
\begin{equation}
 E^s=\left(1+\frac{1}{2}h\right)m^2,
\end{equation}
so that 
\begin{equation}
D_s=\eta^{\mu\nu}\nabla^s_\mu \partial_\nu-E_s.
\end{equation}
One can build the effective curvature tensor from the operator of eq. \ref{eq:nablas} as
\begin{equation}
[\nabla^s_\mu,\nabla^s_\nu]v_\beta=(R^s)^\alpha_{\ \beta \mu\nu}v_\alpha,
\end{equation}
whose invariant contractions define the $A_n$ coefficients.
Each $A_n$ is an invariant under the original symmetries built from a total $n$ powers of $(R^s)^\alpha_{\ \beta \mu\nu}$ and/or $E_s$ \cite{Avramidi:2000bm}
\begin{equation}
A_n=A_n\Big(\left((R^s)^\alpha_{\ \beta \mu\nu}\right)^{n_r},(E^s)^{n_e}\Big),\ \ \ {\rm with} \ \ \ n_r+n_e=n,
\end{equation}
with positive integers $n_r,n_e,n$.
The leading contributions are \cite{Vassilevich:2003xt}
\begin{equation}
A_0=I,\ \ \ A_1=k_1\ \eta^{\mu\nu}(R^s)^\alpha_{\ \mu \alpha\nu}+k_1'E_s,
\end{equation}
where the $k_1,k_1'$ are constants that can be explicitly calculated, but not relevant for this work \cite{Avramidi:2000bm}.

The leading first loop contributions are
\begin{equation}
S'^{(1)}(h)=\int d^4x\ \mathcal{L}'^{(1)}(h)+\mathcal{O}\Big((\partial h)^4\Big),
\end{equation}
with
\begin{equation} \mathcal{L}'^{(1)}(h)=\frac{\Lambda^4}{2}h+
\frac{M_P^2}{2}\left(\frac{1}{2}\partial^\alpha h^{\mu\nu}\partial_\alpha h_{\mu\nu}-\partial_\nu h^{\mu\nu} \partial^\alpha h_{\mu\alpha}+\partial_\alpha h \partial_\mu h^{\mu\alpha}-\frac{1}{2}\partial_\alpha h\partial^\alpha h\right),
\label{eq:dyn}
\end{equation}
where the $M_P$ is the effective Planck scale and $\Lambda$ the cosmological constant. They arise from the integration of $\tau$, the mass of the scalar and the constants $k_1,k_1'$.

The same result can be obtained by adding higher order terms that comply with the original symmetry of the lagrangian. In this case it is the diffeomorphism symmetry, defined by the transformations in eq. \ref{eq:difft} \cite{Blas:2007pp,Alvarez:2019gwz}. 
The second term from eq. \ref{eq:dyn} is the usual dynamical term for a spin 2 field with diffeomorphism symmetry, i.e. the standard graviton \cite{tHooft:1974toh}. 

The dynamical terms are generated and the coupling constants are fixed too. They are defined in terms of the constants of the original theory \cite{Visser:2002ew}.
The masses of the fields, together with the regulator (due to the $\tau$ integral) define the effective coupling constants \cite{Carone:2017mdw,Novozhilov:1991nm,Carone:2018ynf,Carone:2019xot}. In models with scale invariance, these constants are finite without relating to a renormalization scale and defined by the EMT of the theory \cite{Adler:1980pg,Zee:1980sj,Adler:1982ri,Frolov:1997up}. In this setup, the $\mathcal{T}_{\mu\nu}^0$ would define them. The specific dependence of the induced $M_P$ and $\Lambda$ from the constants of the original theory lies beyond the scope of this paper, and they don't change the conclusions. Regardless of their explicit value, it has been shown that  with the FOSEC at one loop, there emerges a calculable effective action for quantum gravity \cite{Buchbinder:1992rb,Kiefer:2012boa}. 

\subsection{Fermion field}
In the Dirac fermion case, the lagrangian in eq. \ref{eq:fosecdir} is already on the form
\begin{equation}
S'(\psi,h)=\int d^4 x\ \bar{\psi}D_\psi \psi+\mathcal{T}^0_{\mu\nu}h^{\mu\nu}+h.c.,
\end{equation}
with
\begin{equation}
D_\psi=\frac{i}{2}[\eta^{\mu\nu}(1+h)-2h^{\mu\nu}]\gamma_\mu\partial_\nu-m(1+h).
\end{equation}
Using the standard background field method for the fermion field, the first loop contribution is 
\begin{equation}
S'^{(1)}(h)=i {\rm Tr} \ln (D_\psi/D_\psi^0)=\frac{i}{2} {\rm Tr} \ln (D_\psi/D_\psi^0)^2,
\label{eq:halfer}
\end{equation}
where $D_\psi^0=D_\phi|_{h^{\mu\nu}}=0$ absorbs the infinities and make the term inside the logarithm dimensionless. The operator is then squared to remove the Dirac matrix dependence. 

The mass 
dependent term does not have any gamma matrices and can be calculated explicitly, with the same result as in the scalar case.
Therefore one can work out the $m=0$ case, knowing that at the end there is a term $\Lambda h$ added in the first loop contribution. Using $\{\gamma_\mu,\gamma_\nu\}=2\eta_{\mu\nu}\mathbb{I}$, the squared operator is
\begin{equation}
D^2_\psi=-\frac{1}{4}\left[\eta^{\mu\nu}(1+2h)-4h^{\mu\nu}\right]\partial_\mu\partial_\nu-\frac{1}{4}\left[2\eta^{\mu\nu}\partial_\mu h-4\partial_\mu h^\mu\nu\right]\partial_\nu + \mathcal{O}(h^3),
\end{equation}
up to second order in $h^{\mu\nu}$. One can obtain the operator
\begin{equation}
\nabla^\psi_\mu v_\nu=\left(\left[\delta^\alpha_\nu\left(\frac{1}{4}+\frac{h}{2}\right)-\eta_{\beta\nu}h^{\beta\alpha}\right]\partial_\mu-\delta^\alpha_\nu\partial_\mu\frac{h}{2}+\eta_{\beta\nu}\partial_\mu h^{\beta\alpha} \right)v_\alpha+ \mathcal{O}(h^3),
\label{eq:nablad}
\end{equation}
and define an effective curvature tensor
\begin{equation}
[\nabla^\psi_\mu,\nabla^\psi_\nu]v_\beta=(R^\psi)^\alpha_{\ \beta \mu\nu}v_\alpha=(R^s)^\alpha_{\ \beta \mu\nu}v_\alpha+ \mathcal{O}(h^3).
\end{equation}
At leading order one can obtain the same result as in eq. \ref{eq:dyn}. The constants $k_1,k_1'$ would have an extra $1/2$ factor, due to the squaring in eq. \ref{eq:halfer} \cite{Broda:2008jr}.

\subsection{Internal global symmetry}
The case with internal global symmetries described in section \ref{sec:gener}, would also induce dynamical terms at the first loop for the emergent fields. If one has a a scalar, the first loop contribution is
\begin{equation}
S'^{(1)}(A)=i{\rm Tr}\ln( D_A/D^0_A), \ \ \ {\rm with} \ \ \ D_A=\eta^{\mu\nu} (\partial_\mu -i\tilde{g}A_\mu)\partial_\nu.
\end{equation}
From it, one can define the differential operator (acting on a vector $v_\nu$) and the curvature tensor
\begin{equation}
\nabla_\mu^Av_\nu=(\partial_\mu -i\tilde{g}A_\mu)v_\nu,\ \ \ [\nabla^A_\mu,\nabla^A_\mu]v_\alpha =F_{\mu\nu} v_\alpha.
\end{equation}
Following the same procedure as in sec. \ref{sec:id}, one obtains  trivial $A_{0,1}$ terms, so that the first contribution is
\begin{equation}
A_2=k_2 \eta^{\mu\alpha}\eta^{\nu\beta}F_{\mu\nu}F_{\alpha\beta},
\end{equation}
that generates the action, at first loop order
\begin{equation}
S'^{(1)}(A)=\int d^4x\ \frac{1}{4}  F^{\mu\nu}F_{\mu\nu}+\mathcal{O}(F^4),
\end{equation}
which is the usual dynamical term for a gauge theory. The same happens with a fermion up to different constants \cite{Broda:2008jr}. 

Indeed it has already been pointed out that strong interactions may arise if the color currents are fixed to vanish in the way that has been described above \cite{Amati:1974rm}. 
One can have an induced gauge theory from fixing the conserved current through quantum processes and their degrees of freedom come from the ones of the original fields \cite{Bjorken:1963vg,Terazawa:1976cx,Terazawa:1976xx}.

\section{Poincar\`e gauge theory}
\label{sec:poincare}

Imposing the FOSEC into a lagrangian is equivalent to adding a graviton whose dynamics arise through quantum corrections. A linear theory is obtained in general.
Self couplings arise from higher order corrections \cite{Padmanabhan:2004xk,Deser:1969wk,Boulware:1974sr}. It has been shown that emergent gravitons have self-couplings consistent with GR \cite{Carone:2017mdw}. In special cases, the emergence of GR can be shown explicitly from the original theory. These cases are the scalar without a potential and the massless fermions.
 
\subsection{Massless scalar}

The lagrangian of a massless is
\begin{equation}
\mathcal{L}(\phi)=\eta^{\mu\nu}\left(\partial_\mu \phi\right)^\dagger \partial_\nu \phi.
\end{equation}
With the FOSEC it becomes
\begin{equation}\begin{split}
\mathcal{L}'(\phi,h)&=\left(\eta^{\mu\nu}\left(1+\frac{1}{2}h\right)-h^{\mu\nu}\right)\left(\partial_\mu\phi\right)^\dagger \partial_\nu\phi+\frac{1}{2}h^{\mu\nu}\mathcal{T}^0_{\mu\nu}.
\end{split}\end{equation}
One can make he substitution
\begin{equation}
\left(\eta^{\mu\nu}\left(1+\frac{1}{2}h\right)-h^{\mu\nu}\right)=\sqrt{-g}g^{\mu\nu},
\end{equation}
which changes the measure as
\begin{equation}
\mathcal{D} h=\mathcal{D}{g}\left(\prod_x \det\frac{\delta h}{\delta g}\right)= \mathcal{D}{g}\left(\prod_x\sqrt{|g|}\right)=\mathcal{D}{g}\ \left(e^{i{\rm Tr}\left( -\frac{i}{2}\ln |g|\right)}\right),
\end{equation}
where determinants of constant matrices can be ignored.
This transforms the FOSEC action as
\begin{equation}
S'(\phi,g)=\left(\int d^4 x\ \mathcal{L}'(\phi,g)\right)-\frac{i}{2}\ {\rm Tr} \ln |g|, 
\label{eq:sg}
\end{equation}
with
\begin{equation}
\mathcal{L}'(\phi,g)=\sqrt{-g}\ g^{\mu\nu}\left(\partial_\mu\phi\right)^\dagger \partial_\nu\phi+\frac{\sqrt{-g}}{2}\left(-\frac{1}{2} g^{\mu\nu}\mathcal{T}_{\mu\nu}^0+\frac{1}{2}\mathcal{T}^0\right),
\label{eq:scg}
\end{equation}
where the first term is equivalent to the lagrangian for the scalar in curved spacetime. Scalars have no intrinsic angular momentum so the lagrangian does not involve any connection. 
The whole lagrangian is fully covariant when $\mathcal{T}^0_{\mu\nu}=\Lambda_0 g_{\mu\nu}$, with $\Lambda_0$ a constant.
The $\ln |g|$ term from the measure change will be discussed in the next section.

\subsubsection{Induced dynamics from massless scalars}
\label{sec:gsid}

One can rewrite the lagrangian from eq. \ref{eq:scg} as
\begin{equation}\begin{split}
\mathcal{L}'(\phi,g)=\sqrt{-g}\phi^\dagger D_{gs}\phi
+\frac{\sqrt{-g}}{2}\left(-\frac{1}{2} g^{\mu\nu}\mathcal{T}_{\mu\nu}^0+\frac{1}{2}\mathcal{T}^0\right),\\
{\rm with}\ \ \ D_{gs}=g^{\mu\nu}\partial_\mu\partial_\nu+\frac{1}{\sqrt{-g}}\partial_\mu(\sqrt{-g}g^{\mu\nu})=(g^{\mu\nu}\partial_\mu-g^{\alpha\beta}\Gamma^{\nu}_{\ \alpha\beta})\partial_\nu,
\label{eq:scgp}
\end{split} \end{equation}
and $\Gamma_{\nu\alpha\beta}=\partial_\alpha g_{\beta\nu}+\partial_\beta g_{\alpha\nu}-\partial_\nu g_{\alpha\beta}$, is the standard Levi-Civita connection.

Following the same procedure as in section \ref{sec:id}, the first loop order contribution is
\begin{equation}
S'^{(1)}(h)=i{\rm Tr}\ln(  \sqrt{-g}  D_{gs}/D^0_s).
\label{eq:s1noln}
\end{equation}
Since they have the same form, one can add the extra term from the change of measure in eq. \ref{eq:sg} to the first loop contribution
\begin{equation}
S'^{(1)}(h)-\frac{i}{2} \ {\rm Tr}\ln |g|=i{\rm Tr}\ln( D_{gs}/D^0_s),
\end{equation}
which is the one loop contribution of a scalar in a curved background without dynamics \cite{Visser:2002ew,Chaichian:2018xdo}. Following the same procedure as in sec. \ref{sec:id}, one can define the operator
\begin{equation}
\nabla^{sg}_\mu v_\nu=(\delta^\alpha_\nu \partial_\mu-\Gamma^\alpha_{\mu\nu})v_\alpha,
\label{eq:dsg}
\end{equation}
which is the standard covariant derivative and defines $D_{gs}=g^{\mu\nu}(\nabla^{gs})_\mu \partial_\nu$. With the standard covariant derivative one can define the Riemann curvature tensor
\begin{equation}
[\nabla^{sg}_\mu,\nabla^{sg}_\nu]v_\alpha=R^\beta_{\ \alpha\mu\nu}v_\beta.
\end{equation}
The coefficients $A_n$ are built of invariants made from contractions of $n$ powers the curvature tensor and the metric tensor
\begin{equation}
A_n=A_n\Big((R^\beta_{\ \alpha\mu\nu})^n,g_{\delta\gamma}\Big).
\end{equation}
The first two leading contributions are \cite{Avramidi:2001ns,Avramidi:2009tn,Chaichian:2018xdo,Azri:2018qcc}
\begin{equation}
A_0=k_0\sqrt{-g},\ \ \ A_1=k_1\sqrt{-g}\ g^{\mu\nu}R^\beta_{\ \mu\beta\nu}=k_1\sqrt{-g} R,
\label{eq:r}
\end{equation}
where $R$ is the Ricci scalar.
Since the Ricci scalar is defined only in terms of $g^{\mu\nu}$, there is no torsion nor nonmetricity generated. 

It is important to remark the physical changes the $\ln|g|$ term makes. In the case without the $\ln|g|$ term,  the coefficients would be calculated from 
eq. \ref{eq:s1noln} directly,  in terms of
\begin{equation}
[\sqrt{-g}\nabla^{sg}_\mu,\sqrt{-g}\nabla^{sg}_\nu]v_\alpha=-g\ R^\beta_{\ \alpha\mu\nu}v_\beta,
\end{equation}
since $\nabla^{sg}_\mu\sqrt{|g|}=0$. While this does not affect $A_0$, it changes the dependence of 
\begin{equation}
A_1=A\Big(|g|R^\beta_{\ \alpha\mu\nu}v_\beta,g_{\delta\gamma}\Big)=\tilde{k}_1\sqrt{-g} R,
\end{equation}
which is the same invariant but with a different constant. Since the functional dependence of the $A_1$ of $R$ is different, the equations it fulfills would change the coefficient \cite{Avramidi:2000bm}. Therefore the effect of the $\ln|g|$ coming from the change of measure can be absorbed into the heat kernel expansion in terms of the same invariants but changing the coefficients. Here they are not calculated explicitly, therefore these changes are not relevant for this work.

At leading order, the first loop contributions are
\begin{equation}
S'^{(1)}(g)-\frac{i}{2}{\rm Tr}\ \ln |g|=\int d^4x\ \mathcal{L}'^{(1)}(g)+\mathcal{O}(R^2),
\end{equation}
with
\begin{equation} \mathcal{L}'^{(1)}(g)=\frac{\Lambda^4}{2}\sqrt{-g}+
\frac{M_P^2}{2}\sqrt{-g} R,
\label{eq:dyngs}
\end{equation}
where the effective cosmological constant and Planck scale are obtained in terms of the $k$ coefficients and the regularization of the $\tau$ integration from eq. \ref{eq:tau}, together with the value of $\mathcal{T}^0_{\mu\nu}$.

In the case where $\mathcal{T}^0_{\mu\nu}=\Lambda_0 g_{\mu\nu}$, the resulting theory has full coordinate invariance.

The standard lagrangian for a massless scalar with GR and quantized metric has been obtained.
One can conclude that, in the massless scalar case, one obtains standard GR at leading order.

\subsection{Massless fermions}

In section  \ref{sec:globloc}, it was shown that a gauge theory, up to linear order, can be obtained from a theory with global symmetry. This is done by requiring the off-shell vanishing of its conserved current. In the case of a fermion lagrangian, the gauge coupling is linear so the full gauge theory is obtained.

When considering fermions, the FOSEC should be enhanced with the requirement of a fixed angular momentum tensor. It is independent of the EMT, due to the non trivial Lorentz representation of the fermions. This would generate the full $T^4\ltimes SO(3,1)$ Poincar\`e gauge theory which has GR as its subset \cite{Obukhov:2018bmf,Blagojevic:2013xpa,CordeirodosSantos:2019ngd}.

The lagrangian of a massless fermion $\psi$ is
\begin{equation}
\mathcal{L}(\psi)=\frac{i}{2}\bar{\psi}\delta_a^\mu\gamma^a\partial_\mu\psi+h.c.,
\end{equation}
where the spacetime indices $\mu$ are different from the spin ones $a$. Imposing the FOSEC with the EMT
\begin{equation}
T_{\mu}^a(\psi)=\frac{i}{2}\bar{\psi}\gamma^{a}\partial_{\mu}\psi-\frac{i}{2}\delta^a_\mu\delta^\nu_b\bar{\psi}\gamma^b\partial_\nu\psi+h.c.,
\end{equation}
generates the lagrangian
\begin{equation}
\mathcal{L}'(\psi,e)=\frac{i}{2}\left(\delta^\mu_a\left(1+\delta_\nu^b e_b^\nu\right)+e^{\mu}_a\right)\bar{\psi}\gamma^a\partial_\mu\psi+ e_a^\mu \mathcal{T}_\mu^{0a}+h.c.
\label{eq:psie}
\end{equation}
The angular momentum tensor has an orbital part, which is fixed off-shell if the EMT is. The spin tensor is independent from the EMT and it can be required to be fixed off-shell too. It will be required that both the EMT and the spin tensor be fixed off-shell simultaneously. This is done by fixing the spin tensor of the FOSEC fulfilling lagrangian from eq. \ref{eq:psie}, which is
\begin{equation}
S^{\mu bc}=\left(\delta^\mu_a\left(1+\delta_\nu^b e_b^\nu\right)+e^{\mu}_a\right) \bar{\psi}\gamma^a\sigma^{bc}\psi,\ \ \ {\rm where} \ \ \ \sigma^{ab}=\frac{i}{2}[\gamma^a,\gamma^b].
\end{equation}
Requiring this tensor to vanish off-shell yields the lagrangian
\begin{equation}
\mathcal{L}'(\psi,e,\omega)=\frac{i}{2}\left(\delta^\mu_a\left(1+\delta_\nu^b e_b^\nu\right)+e^{\mu}_a\right)\bar{\psi}\gamma^a\left(\partial_\mu-\frac{i}{4}\omega_{\mu bc}\sigma^{bc}\right)\psi+e^\mu_a\mathcal{T}_{\mu}^{0a}+h.c.,
\end{equation}
where $\omega_{\mu ab}$ is the spin connection, whose integration fixes $S^{\mu a b}$ to vanish. It is multiplied by an arbitrary constant chosen to be $i/4$ to resemble the spin connection in curved spacetime. The $e^\mu_a$ can be redefined as
\begin{equation}
\tilde e^\mu_a=e^{\mu}_a+\delta^\mu_a\left(1+\delta_\nu^b e_b^\nu\right),
\end{equation}
which doesn't change the measure, up to an irrelevant constant. The lagrangian can be rewritten as
\begin{equation}
\mathcal{L}'(\psi,\tilde e,\omega)=\frac{i}{2}\tilde e^{\mu}_a\bar{\psi}\gamma^a\left(\partial_\mu-\frac{i}{4}\omega_{\mu bc}\sigma^{bc}\right)\psi+\tilde  e^\nu_b(\delta^\mu_\nu\delta^b_a-\delta_\nu^b\delta^\mu_a/5)\mathcal{T}_{\mu}^{a0}+h.c.,
\end{equation}
which is already equivalent to the curved space fermion lagrangian with the Goldberg variables \cite{Scharf:2001bk}
\begin{equation}
\tilde e^\mu_a=|e| e^\mu_a.
\end{equation} 
This changes the measure as
\begin{equation}
\mathcal{D} \tilde e=\mathcal{D}e\left(\prod_x \det\frac{\delta \tilde{e}}{\delta e}\right)= \mathcal{D} e\left(\prod_x |e|\right)=\mathcal{D}e\left(e^{i{\rm Tr}\left( -i\ln |e|\right)}\right),
\end{equation}
where determinants of constant matrices can be ignored.
This transforms the FOSEC action
\begin{equation}
S'(\psi,e,\omega)=\left(\int d^4 x \mathcal{L}'(\psi,e,\omega)\right)-i\ {\rm Tr} \ln |e|, 
\label{eq:dg}
\end{equation}
with the lagrangian 
\begin{equation}\begin{split}
\mathcal{L}'(\psi, e,\omega)&=\frac{i}{2}|e| e^{\mu}_a\bar{\psi}\gamma^a\left(\partial_\mu-\frac{i}{4}\omega_{\mu bc}\sigma^{bc}\right)\psi\\
&\quad+|e|  e^\nu_b(\delta^\mu_\nu\delta^b_a-\delta_\nu^b\delta^\mu_a/5)\mathcal{T}_{\mu}^{a0}+h.c.,
\label{eq:fergr}
\end{split}\end{equation}
whose first line is the standard fermion lagrangian in curved space as in eq. \ref{eq:curvfer} with no dynamical terms for $e^{\mu}_a,\omega_{\mu ab}$. The tetrad field is independent of the spin connection. This is in general true since the spin connection fully contains the information of the torsion and nonmetricity, which are not related to the metric. In this emergent setup, the spin connection arises from fixing the spin tensor, which is antisymmetric. There is no reason for the symmetric part of the spin connection to exist. Therefore no nonmetricity can be generated, but torsion can be nonzero. The existence of an independent gauge symmetry would forbid the appearance of torsion.

\subsubsection{Induced dynamics from massless fermions}

One can follow the same procedure as in sec. \ref{sec:id}. 
By writing the FOSEC lagrangian from \ref{eq:fergr} as
\begin{equation}
\mathcal{L}'(\psi, e,\omega)=\frac{i}{2}|e| \bar{\psi}D_{g\psi}\psi
+|e|  e^\nu_b(\delta^\mu_\nu\delta^b_a-\delta_\nu^b\delta^\mu_a/5)\mathcal{T}_{\mu}^{a0}+h.c.,
\end{equation}
with the operator
\begin{equation}
D_{\psi g}=\frac{i}{2}|e| e^\mu_a \gamma^a\left(\partial_\mu-\frac{i}{4}\omega_{\mu bc}\sigma^{bc}\right),
\end{equation}
one obtains the first loop contributions 
\begin{equation}
S'^{(1)}(e)-i\ {\rm Tr}\ln |e|=i{\rm Tr}\ln( D_{g\psi}/D^0_\psi)=\frac{i}{2}{\rm Tr}\ln( D_{g\psi}/D^0_\psi)^2.
\end{equation}
The absorption of the $\ln|e|$ term works just as in sec. \ref{sec:gsid}.
The differential operator is squared to eliminate the gamma matrix dependence. After some simplifications it becomes \cite{Shapiro:2016pfm}
\begin{equation}
(D_{\psi g})^2=\left(g^{\mu\nu}\partial_\mu-g^{\alpha\beta}\Gamma^\nu_{\ \alpha\beta}-\frac{i}{4}g^{\mu\nu}\omega_{\mu b c}\sigma^{bc}\right)\left(\partial_\nu-\frac{i}{4}\omega_{\nu fg}\sigma^{fg}\right),
\end{equation}
where the $\Gamma^\nu_{\ \alpha\beta}$ is the standard Christoffel symbol. One can obtain the exact same differential operator acting on a vector as in eq. \ref{eq:dsg}, and the differential operator acting on fermions
\begin{equation}
\nabla^{\psi g}_\mu \Psi=\left(\partial_\mu-\frac{i}{4}\omega_{\mu a b}\sigma^{ab}\right)\Psi.
\end{equation}
Note that the spin connection is unrelated to the metric tensor. One can obtain the invariant terms from these operators 
\begin{equation}
[\nabla^{sg}_\mu,\nabla^{sg}_\nu]v_\alpha=R^\beta_{\ \alpha\mu\nu}v_\beta,\ \ \ [\nabla^{\psi g}_\mu,\nabla^{\psi g}_\nu]\Psi=\Omega_{\mu\nu a b}\sigma^{ab}\Psi, 
\end{equation}
where 
\begin{equation}
\Omega_{\mu\nu}^{\ \  a b}=\partial_{[\mu} \omega_{\nu]}^{\ \ ab}+\omega_{[\mu}^{\ \ ac}\omega_{\nu]c}^{\ \ \ b}.
\end{equation}
Since they are independent, the leading order terms in the expansion are a sum of the independent invariants
 \cite{Avramidi:2009tn,Chaichian:2018xdo}
\begin{equation}
A_0=k_0 |e|,\ \ \ A_1= k_1 |e| R+ k'_1 |e|e^\mu_a e^\nu_b \Omega_{\mu\nu}^{\ \  a b}.
\end{equation}
The first loop contributions are
\begin{equation}
S'^{(1)}(e)-i{\rm Tr}\ \ln |e|=\int d^4x\ \mathcal{L}'^{(1)}(e)+\mathcal{O}(R^2+\Omega^2),
\end{equation}
with
\begin{equation} 
 \mathcal{L}'^{(1)}(e)=|e|\frac{\Lambda^4}{2}+   \frac{M_P^2}{2(k_1+k'_1)} |e|(k_1 R +k'_1e^\mu_a e^\nu_b \Omega_{\mu\nu}^{\ \  a b}).
\label{eq:dyngp}
\end{equation}
 At this order, the equations of motion of the spin connection are
\begin{equation}
\partial_{[\mu}e_{\nu}^a+\omega_{[\mu\ b}^{\ \ a}e_{\nu]}^{b}=0,
\label{eq:pala}
\end{equation}
relating the spin connection to the metric \`a la Palatini. However this relation happens only at first order. Since there are higher order terms, the actual relation would be
\begin{equation}
\omega_{\mu a b}=\hat{\omega}_{\mu ab}+\mathfrak{w}_{\mu ab},
\end{equation}
where $\hat{\omega}_{\mu ab}$ satisfies eq. \ref{eq:pala} and $\mathfrak{w}_{\mu ab}$ is related to torsion and nonmetricity, whose invariant dynamical terms are also induced and are included in the terms of eq. \ref{eq:dyngp}
\cite{BeltranJimenez:2019tjy,Chaichian:2018xdo}. 
Since the object arises as a Lagrange multiplier of the spin tensor it must be antisymmetric. Therefore the nonmetricity is zero \cite{Adak:2006rx}. If there where to be a gauge symmetry, the appearance of torsion would be forbidden, then eq. \ref{eq:pala} would be exact.

In this case GR has emerged enhanced (with a possible nonzero torsion when there are no gauge symmetries) and zero nonmetricity. If $\mathcal{T}^0_{\mu\nu}=\Lambda_0 g_{\mu\nu}$ and the spin tensor vanishes, the model is fully covariant.

It was shown that a lagrangian with a massless fermion that fulfills the FOSEC, together with the off-shell vanishing of the spin tensor, generates the full Poincar\`e gauge theory which has GR as a subset.  The standard lagrangian for a massless fermion in GR (when torsion vanishes) with quantized tetrad field has been obtained.

\section{Emergent gravity}
\label{sec:eg}

Assuming the FOSEC on a lagrangian implies the addition of linear gravity to that lagrangian. In the special cases shown in sec. \ref{sec:poincare} the complete Poincar\`e gauge theory or GR appear \cite{Obukhov:2018bmf,Aldrovandi:2013wha,Hehl:2019csx}.
In these specific cases, the dynamics of the fields fulfilling the FOSEC resemble the dynamics of the fields in curved spacetime, i.e. under the influence of gravity. Therefore gravity-like effects emerge from imposing the FOSEC. 

The obtained gravitational theory is an effective theory, and it contains an infinite amount of higher order terms besides the Ricci scalar $R$ (that generates GR). All the gravitational coupling constants are induced from the original theory. 
The explicit calculation of the constants is a standard procedure and lies beyond the scope of this paper. One of the main problems is that their specific value is dependent of the regulator \cite{Novozhilov:1991nm}.
In specific theories, such as scale invariant ones, the gravitational constants are calculable from the ones of the original theory (incluiding $\mathcal{T}_{\mu\nu}^0$) \cite{Adler:1982ri,Novozhilov:1991nm}. This would solve the problem that the quantization of GR requires infinite arbitrary coupling constants (i.e. non renormalizable), by relating all of them to the finite number of original coupling constants and $\mathcal{T}_{\mu\nu}^0$. As it has been mentioned above, the requirement of full covariance fixes $\mathcal{T}_{\mu\nu}^0=\Lambda_0 g_{\mu\nu}$, so that $\Lambda_0$ defines the gravitational interaction scale.  Higher order terms appear from higher order loop contributions, therefore they are naturally suppressed and avoid any inherent problems \cite{Donoghue:1994dn}.

An usual problem for adding spin 2 particles into a model is that they must comply with Weinberg-Witten theorem. It states that, if there are massless spin 2 states $|p\rangle,|p'\rangle$, then
\begin{equation}
\lim_{p'\to p} \langle p'| T_{\mu\nu}^{h} | p\rangle=0,
\end{equation}
where $T_{\mu\nu}^{h}$ is the EMT of the graviton. This forbids non graviton spin 2 particles \cite{Jenkins:2009un,Loebbert:2008zz}.
This theorem does not apply to the graviton emerging from the FOSEC. When matter is on-shell, the EMT is already conserved and no emergent gravitons are necessary.  The emergent gravitons are auxiliary fields with no on-shell states. They appear to fulfill the FOSEC, just as ghost appear when fixing the gauge. Just like ghosts, they should appear in calculations through their propagator and couplings but they don't appear as external legs.
The theorem is avoided in this setup since there are no on-shell $|p\rangle$ states for the emergent graviton.

The dynamical terms are generated as in Sakharov's induced gravity, but lacking most of its usual problems \cite{Chaichian:2018xdo}. The main one being the assumption of an arbitrary unknown curved spacetime without dynamics, without a physical reason. Here, the actual spacetime is always flat but the dynamics of the original fields change (through the FOSEC) so that they behave as if they were on a curved spacetime. Even though the mathematical structure is the same as in curved spacetime, the emergent metric is not related to the spacetime curvature but to a change in the behavior of the off-shell field configurations that fulfill the FOSEC. The FOSEC implies curvature to emerge in the quantum vacuum and not in the actual spacetime. Therefore this curvature emerges from a  purely quantum process, reminiscent of the 
ER=EPR idea \cite{Maldacena:2013xja}.

One can also reverse engineer the ideas in this work. One can assume a QFT with a quantized metric for a curved spacetime, then forbid any non-renormalizable  kinetic terms. The metric without dynamics can be integrated out This imposes the FOSEC.

An emergent theory of gravity has been obtained. The underlying theory lies in a flat spacetime. Imposing the FOSEC changes the quantum vacuum by limiting the available off-shell configurations. In general a linear theory is obtained, while in specific models a non linear GR-like theory can emerge.
 The FOSEC creates gravity-like effects resembling a quantized metric of a curved spacetime \cite{Carlip:2012wa}. The emergent nature of the dynamical terms fix the effective couplings in terms of the original ones.

\section{Conclusions}

It has been shown that imposing the FOSEC on a general lagrangian implies the appearance of a graviton without dynamics. The dynamical terms are induced \`a la Sakharov through quantum corrections. This has been explicitly shown for a lagrangian with a complex scalar and a fermion. In general the FOSEC implies the emergence of linear quantum gravity.

In the case of a massless fermion, the FOSEC, together with the off-shell vanishing of the spin tensor requirement, implies the appearance of the full Poincar\`e gauge theory, with GR as a subset. In the case of scalar field without a potential, GR can be obtained. Both cases appear with a quantized metric. All dynamical terms are obtained through quantum corrections. This induces gravity \`a la Sakharov but does not need any prior assumption on the spacetime geometry, that always stays flat. The emergent metric arises from the changes of the off-shell configurations, when complying with the FOSEC. Although the metric is mathematically equivalent to the one that comes from curved spacetime, conceptually it originates from a modification in the quantum behavior of the original fields. 
This way, a quantized graviton emerges, whose non-renormalizable couplings are all effectively calculable from the 
original theory and the EMT of the original system. This evades the Weinberg-Witten theorem, since the emergent gravitons are auxiliary fields without off-shell configurations. 

It was shown that imposing the FOSEC on a lagrangian, causes a quantized gravity-like metric field to emerge in that lagrangian. The spacetime is always flat and imposing the FOSEC implies the emergence of a quantized metric whose curvature describes modifications on the quantum vacuum that drive off-shell processes. Therefore imposing the FOSEC implies an emergent quantum gravity.

\end{document}